\documentclass[a4paper]{article}
\usepackage{geometry}
\usepackage{amsmath}
\usepackage{amsfonts}
\usepackage{amsmath,amssymb}
\usepackage{graphicx}
\usepackage{color}
\usepackage{threeparttable}

\renewcommand{\baselinestretch}{1}
\renewcommand{\arraystretch}{1.3}

\newtheorem{theorem}{Theorem}
\newtheorem{lemma}{Lemma}

\newtheorem{definition}{Definition}

\newcommand{\ls}[1]
    {\dimen0=\fontdimen6\the\font\lineskip=#1\dimen0
     \advance\lineskip.5\fontdimen5\the\font
     \advance\lineskip-\dimen0
     \lineskiplimit=0.9\lineskip
     \baselineskip=\lineskip
     \advance\baselineskip\dimen0
     \normallineskip\lineskip\normallineskiplimit\lineskiplimit
     \normalbaselineskip\baselineskip
     \ignorespaces}

\begin{document}

\bibliographystyle{abbrv}

\title{Double Circulant Self-dual Codes on Sextic Cyclotomy}
\author{Tongjiang Yan$^{\mathrm{a}}$,  Tao Wang$^{\mathrm{a}}$,  Wenpeng Gao$^{\mathrm{a}}$, Xvbo Zhao$^{\mathrm{a}}$\\
$^a$ College of Science,
China University of Petroleum,\\
Qingdao 266555,
Shandong, China\\
Corresponding author: Tongjiang Yan\\
Email: yantoji@163.com; 631941071@qq.com; \\
wenpenggao@foxmail.com; zhaoxubo$\_$2003@163.com}
\maketitle
\thispagestyle{plain} \setcounter{page}{1}
\begin{abstract}
This paper contributes to construct double circulant self-dual codes by sextic  cyclotomy. Generator matrixes of a  family of  pure double circulant codes and a  family of double circulant codes
with boundary are formed from sextic  cyclotomic classes. These codes are proved to be self-dual on certain conditions. Moreover, experiments show that some of them are optimal self-dual codes or best than ever codes over $\mathrm{GF}(2)$ and $\mathrm{GF}(4)$.



\end{abstract}

\ls{1.2}
\section{Introduction}\label{section 1}

An $[n,k,d]$ linear code $C$ over the Galois field $\mathrm{GF}(q)$ is a $k$-dimensional subspace of  $\mathrm{GF}(q)^{n}$ with the minimum Hamming weight $d$, where $q$ is a prime power. Its generator matrix $G$ is a $k \times n$ matrix over $\mathrm{GF}(q)$ whose rows span the code. Every element of $C$ is called a codeword. The Euclidean inner product between two codewords $x = (x_0, x_1, \dots, x_{n-1})$ and $y = (y_0, y_1, \dots, y_{n-1})$ is defined by
\begin{center}
$(x, y)=\sum\limits_{i=0}^{n-1}x_{i}y_{i}.$
\end{center}
The orthogonal subspace of $C$  in the linear space $\mathrm{GF}(q)^{n}$ denoted by
\begin{center}
$C^\perp=\{x\in \mathrm{GF}(q)^n|(x,c)=0 ,  \textrm{for all } c \in C\}$
\end{center}
is defined as an Euclidean dual code of $C$.  Moreover, the code $C$ is  Euclidean self-dual
if $C = C^\perp$, hereinafter referred to as a self-dual code.

Self-dual codes play an important role in classical linear error correcting codes and constructing quantum error correcting codes \cite{Nebe2006Self, Pless1975On}.  But it is difficult to construct them with larger minimum Hamming distances. Although the existence of progressive self-dual code groups has been proved, we are still unable to find out exactly in most cases. A double circulant self-dual code employs one generator matrix consists  of two circulant matrices and possesses  many good algebraic properties \cite{Karlin1969New,Gaborit2002Quadratic}. With more new double circulant self-dual codes were presented, these  provide a remarkable way to construct self-dual codes \cite{Gaborit2002Quadratic, Tao2015Fourth, T2019On, MinjiaShi1, MinjiaShi2, MinjiaShi3, MinjiaShi4}. Recently, by constructting double circulant generator matrixes using  cyclotomy of order four modulo an odd prime $p$, Zhang  and Ge  presented some double circulant self-dual codes with optimal minimum distances \cite{Tao2015Fourth}.  Next, we will construct several double circulant self-dual codes with  optimal minimum distances by sextic cyclotomy modulo $p$. All computations have been done by MAGMA \cite{Bosma1997The}.

The Hamming weight of a codeword, denoted by $wt(x)$, is defined as the number of  nonzero coordinates of $x$. The minimum nonzero Hamming weight of all codewords in a linear code $C$ is equal to its minimum Hamming distance $d(C)$.

 \begin{lemma}\cite{Nebe2006Self}\label{Lemma 2.1} Let $C$ be a self-dual code.

(1) For the case $q=2$,
\begin{center}
$d(C)\leq
\begin{cases}
4\left\lfloor \displaystyle\frac{n}{24} \right\rfloor+4, & \textrm{$n\not\equiv 22 \bmod{24}$,}\vspace{3mm}\\

4\left\lfloor \displaystyle\frac{n}{24} \right\rfloor+6, & \textrm{$n\equiv 22 \bmod{24}$.}
\end{cases}$
\end{center}

(2) For the case $q=4$,  $$d(C)\leq 4\left\lfloor \displaystyle\frac{n}{12} \right\rfloor+3.$$
\end{lemma}

The self-dual code $C$ is called optimal if and only if it can get the highest possible minimum distance.  If the minimum Hamming distance of the self-dual code exceeds those of all the codes previously constructed, we call it best than ever. All the known optimal and  best than ever self-dual codes can be found in \cite{Gaborit2020Tables}.

This paper is organized as follows.  We firstly present some definitions and preliminaries in Section $2$. In Section $3$, several families of double circulant self-dual codes are obtained. Section $4$ concludes the paper finally.

\section{Primaries}\label{section 2}
\subsection{cyclotomy}
Suppose $p = 6f + 1$ is an odd prime and  $f$ is a positive integer.   Let $\gamma$ be a fixed  primitive element on the Galois field $\mathrm{GF}(p)$.
We define a sextic cyclotomic classes $C_i$ as
\begin{equation*}
C_i = \{\gamma^{6j+i} : 0\leq j \leq \displaystyle\frac{p-1}{6} -1\},
\end{equation*}
where $0 \leq i \leq 5$. That means $C_i = \gamma^{i}C_0$. For arbitrary integers $m, n$ satisfying $0 \leq m,n \leq 5$, a sextic cyclotomic number is define as \cite{Tao2015Fourth} :
\begin{equation*}
(m,n)= \left|C_{m}+1\cap C_{n}\right|.
\end{equation*}

The following lemma summarizes some basic properties of the cyclotomic number.
\begin{lemma}\label{lemma 2} \cite{GJS1998}
Let $p = 6f + 1$ be some odd prime. Then
\begin{equation*}
\begin{split}
&(a)\qquad (i,j) = (i',j'), \;when\;i \equiv i'\bmod {6},\; j \equiv j'\bmod {6}; \\
&(b)\qquad (i,j) = (6-i,j-i) = \left\{
\begin{aligned}
(j,i),\qquad\quad f\; is\; even; \\
(j+3,i+3),\; f\; is \; odd; \\
\end{aligned}
\right. \\
&(c)\qquad \sum\limits_{i=0}^{5}=f-\delta_j,
\;where\; \delta_j=
\left\{
\begin{aligned}	
1, &\; if \;j=0\bmod {6};\\
0, & \;otherwise.
\end{aligned}
\right.
\end{split}
\end{equation*}
\end{lemma}
\begin{lemma}\label{theorem 3}
Suppose $p=12l+7$ is a prime. Then
\begin{equation*}
\begin{aligned}
(0,0)=&(3,0)=(3,3),\,\, (0,1)=(2,5)=(4,3),\,\, (0,2)=(1,4)=(5,3)\,\,\,(0,4)=(1,3)=(5,2)\\
(0,5)=&(2,3)=(4,1),\,\, (1,0)=(2,2)=(3,4)=(4,0)=(3,1)=(5,5),\,\,\,(2,1)=(4,5)\\
(1,1)=&(2,0)=(3,2)=(3,5)=(4,4)=(5,0),\,\,(1,2)=(1,5)=(2,4)=(4,2)=(5,1)=(5,4).
\end{aligned}
\end{equation*}
\end{lemma}
$\mathbf{Proof.}$
The proof is straightforward from the Lemma \ref{lemma 2}. $\blacksquare$

For simplicity, in the following  we define
\begin{equation*}
\begin{aligned}	
A:=&(0,0), \qquad B:=(0,1), \qquad C:=(0,2), \qquad D:=(0,3),  \qquad E:=(0,4),\\
F:=&(0,5),\qquad G:=(1,0), \qquad H:=(1,1), \qquad I:=(1,2), \qquad J:=(2,1).
\end{aligned}
\end{equation*}
\begin{lemma}\label{theorem 4} \cite{GJS1998}
Suppose $p = 12l + 7$ is a prime. Let $g$ be a primitive root of $p$, satisfying $g^t\equiv 2 \bmod{p}$, where $t$ is a positive integer. Then there are positive integers $x$, $y$, such that $p=x^2+3y^2$, and $x \equiv 1 \bmod{3}$, $y \equiv -t\bmod {3}$. Thus the sextic cyclotomic numbers are\\
(a) when $t\equiv0\bmod3$,
\begin{equation*}
\begin{aligned}
A&=\frac{p-11-8x}{36}, \,\, \,\,\,\qquad B=\frac{p+1-2x+12y}{36}, \qquad C=\frac{p+1-2x+12y}{36}, \\
D&=\frac{p+1+16x}{36},  \,\,\,\,\,\qquad E=\frac{p+1-2x-12y}{36}, \,\,\,\,\,\,\quad F=\frac{p+1-2x-12y}{36},\\
G&=\frac{p-5+4x+6y}{36},\quad\ H=\frac{p-5+4x-6y}{36},  \,\,\qquad I=\frac{p+1-2x}{36}, \\
J&=\frac{p+1-2x}{36};
\end{aligned}
\end{equation*}
(b) when $t\equiv1\bmod3$,
\begin{equation*}
\begin{aligned}
A=&\frac{p-11-2x}{36},\,\,\,\, \qquad \qquad B=\frac{p+1+4x}{36}, \qquad \,\, \,\,\qquad C=\frac{p+1-2x+12y}{36}, \\
D=&\frac{p+1+10x-12y}{36},\qquad E=\frac{p+1-8x-12y}{36}, \qquad F=\frac{p+1-2x+12y}{36},\\
G=&\frac{p-5-2x+6y}{36}, \,\,\,\,\,\qquad H=\frac{p-5+4x-6y}{36}, \,\,\,\, \,\qquad I=\frac{p+1+4x}{36},\\
J=&\frac{p+1-8x+12y}{36};
\end{aligned}
\end{equation*}
(c) when $t\equiv2\bmod3$,
\begin{equation*}
\begin{aligned}
A=&\frac{p-11-2x}{36}, \,\,\,\qquad\qquad  B=\frac{p+1-2x-12y}{36},\,\qquad  C=\frac{p+1-8x+12y}{36}, \\
D=&\frac{p+1+10x+12y}{36}, \qquad  E=\frac{p+1-2x-12y}{36}, \qquad F=\frac{p+1+4x}{36},\\
G=&\frac{p-5+4x+6y}{36}, \,\,\,\,\, \qquad H=\frac{p-5-2x-6y}{36}, \,\,\,\,\qquad I=\frac{p+1+4x}{36},\\
J=&\frac{p+1-8x+12y}{36}.
\end{aligned}
\end{equation*}
\end{lemma}
\subsection{Sextic residue double circulant codes}
Let $p=6f+1$ be an odd prime, where $f$ is a positive integer.  Let $\vec{m}:=(m_0,m_1,m_2,m_3,m_4,m_5,m_6) \in \mathrm{GF}(q)^7$, where $q$ is a prime power. Now we define $C_p(\vec{m})$ as a  matrix $R=(r_{ij})_{p\times p}$ on $\mathrm{GF}(q)$, where $1 \leq i,j \leq p$, and

\begin{equation}\label{r}
r_{ij}=\left\{
\begin{aligned}
m_0, &\quad \mathrm{if}\quad j=i,\\
m_{s+1}, & \quad \mathrm{if} \quad j-i \in C_s,
\end{aligned}
\right.
\end{equation}
where $s+1$ means $s+1\bmod{6}$, and $s=0,1,\cdots,5$.

\begin{definition}\label{definition 7}
Let $P_p(R)$ and $B_p(\alpha,R)$ be two linear codes with the following generator matrixes
\begin{equation*}
\left(
\begin{array}{cc}
I_p, R
\end{array}
\right),\,\,\,\,\,\,\,\,
\left(
\begin{array}{cccccc}
\quad  &  \alpha  &  1     & \cdots  & 1 \\
\quad  &  -1      &  \quad & \quad   & \quad \\
I_{p+1}&  \vdots  &  \quad & R       & \quad \\
\quad  &  -1      &  \quad & \quad   & \quad
\end{array}
\right)
\end{equation*}
respectively, where $\alpha \in \mathrm{GF}(q)$ and $R$ is the cyclic matrix defined in Equation (\ref{r}). We call $P_p(R)$ a  sextic residue  pure double circulant code and $B_p(\alpha,R)$  a  sextic residue  double circulant code with boundary. In general, we call them sextic residue double circulant codes.
\end{definition}

Define $A_0$ and $J_p$ as the $p \times p$ identity matrix and all-one matrix, respectively. Then $C_p(1,0,0,0,0,0,0)=A_0$, $C_p(1,1,1,1,1,1,1)=J_p$. Define
\begin{equation}\label{3-10}
\begin{aligned}	
A_1:=&C_p(0,1,0,0,0,0,0) \qquad A_2:=C_p(0,0,1,0,0,0,0) \\
A_3:=&C_p(0,0,0,1,0,0,0) \qquad A_4:=C_p(0,0,0,0,1,0,0) \\
A_5:=&C_p(0,0,0,0,0,1,0) \qquad A_6:=C_p(0,0,0,0,0,0,1) \\
\end{aligned}
\end{equation}
Then we can get the following results.
\begin{lemma}\label{lemma 5}
Suppose $p$ is a prime and $p=12l+7$. Then the matrices in Equation (\ref{3-10}) has the following properties
\begin{equation*}
\begin{aligned}	
A_1=&A_4^T, \qquad A_2=A_5^T, \qquad A_3 = A_6^T, \\
A_1^2=&AA_1+BA_2+CA_3+DA_4+EA_5+FA_6, \\
A_2^2=&FA_1+AA_2+BA_3+CA_4+DA_5+EA_6, \\
A_3^2=&EA_1+FA_2+AA_3+BA_4+CA_5+DA_6, \\
A_4^2=&DA_1+EA_2+FA_3+AA_4+BA_5+CA_6, \\
A_5^2=&CA_1+DA_2+EA_3+FA_4+AA_5+BA_6, \\
A_6^2=&BA_1+CA_2+DA_3+EA_4+FA_5+AA_6,\\	
A_1A_2=&A_2A_1=GA_1+HA_2+IA_3+EA_4+CA_5+IA_6, \\
A_1A_3=&A_3A_1=HA_1+JA_2+GA_3+FA_4+IA_5+BA_6, \\
A_1A_4=&A_4A_1=(2l+1)A_0+AA_1+GA_2+HA_3+AA_4+GA_6+HA_6, \\
A_1A_5=&A_5A_1=GA_1+FA_2+IA_3+BA_4+HA_6+JA_6, \\
A_1A_6=&A_6A_1=HA_1+IA_2+EA_3+CA_4+IA_6+GA_6,\\
A_2A_3=&A_3A_2=IA_1+GA_2+HA_3+IA_4+EA_5+CA_6, \\
A_2A_4=&A_4A_2=BA_1+HA_2+JA_3+GA_4+FA_5+IA_6,\\
A_2A_5=&A_5A_12=(2l+1)A_0+HA_1+AA_2+GA_3+HA_4+AA_5+GA_6, \\
A_2A_6=&A_6A_2=JA_1+GA_2+FA_3+IA_4+BA_5+HA_6, \\
A_3A_4=&A_4A_3=CA_1+IA_2+GA_3+HA_4+IA_5+EA_6, \\
A_3A_5=&A_5A_3=IA_1+BA_2+HA_3+JA_4+GA_5+FA_6, \\
A_3A_6=&A_6A_3=(2l+1)A_0+GA_1+HA_2+AA_3+GA_4+HA_5+AA_6, \\
A_4A_5=&A_5A_4=EA_1+CA_2+IA_3+GA_4+HA_5+IA_6, \\
A_4A_6=&A_6A_4=FA_1+IA_2+BA_3+HA_4+JA_5+GA_6, \\
A_5A_6=&A_6A_5=IA_1+EA_2+CA_3+IA_4+GA_5+HA_6, 
\end{aligned}
\end{equation*}
where $A_i^T$ means the transpose of the matrix $A_i$, $i=0,1,\cdots,6.$

\end{lemma}
$\mathbf{Proof.}$
The proof can be drawn directly from the definitions of $A_i$ and Lemma \ref{theorem 3}. $\blacksquare$
\begin{theorem}\label{theorem 6}
Suppose $p=12l+7=6f+1$ is an odd prime. Then
\begin{equation*}
\begin{aligned}
&\sum_{i=0}^{6}m_iA_i(\sum_{i=0}^{6}m_iA_i)^T=\sum_{i=0}^{6}D_i(\vec{m})A_i,
\end{aligned}	
\end{equation*}
where
\begin{equation*}
\begin{aligned}
D_0(\vec{m})=&m_0^2+f(m_1^2+m_2^2+m_3^2+m_4^2+m_5^2+m_6^2),\\
D_1(\vec{m})=D_4(\vec{m})=&(m_0m_4+m_0m_1)+(m_1^2+m_1m_4+m_4^2)A+(m_1m_2+m_4m_5+m_3m_6)B \\
       &+(m_1m_3+m_4m_6+m_2m_5)C+m_1m_4D+(m_3m_6+m_2m_4+m_1m_5)E \\
	   &+(m_2m_5+m_3m_4+m_1m_6)F+(m_1m_2+m_1m_5+m_2m_4+m_3^2+m_4m_5+m_6^2)G \\
	   &+(m_1m_3+m_1m_6+m_2^2+m_3m_4+m_5^2+m_4m_6)H+(m_2m_6+m_2m_3 \\
       &+m_3m_5+m_3m_5+m_5m_6+m_2m_6)I+(m_2m_3+m_5m_6)J,  \\
D_2(\vec{m})=D_5(\vec{m})=&m_0m_5+m_0m_2+(m_2^2+m_5^2+m_2m_5)A+(m_1m_4+m_2m_3+m_5m_6)B \\
	       &+(m_2m_4+m_1m_5+m_3m_6)C+m_2m_5D+(m_1m_4+m_3m_5+m_2m_6)E \\
	       &+(m_1m_2+m_3m_6+m_4m_5)F +(m_1^2+m_2m_3+m_2m_6+m_4^2+m_3m_5+m_5m_6)G \\
	       &+(m_1m_5+m_1m_2+m_3^2+m_2m_4+m_6^2+m_4m_5)H\\
	       &+(m_1m_3+m_1m_3+m_3m_4+m_4m_6+m_1m_6+m_4m_6)I+(m_1m_6+m_3m_4)J,\\
D_3(\vec{m})=D_6(\vec{m})=&m_0m_6+m_0m_3+(m_3^2+m_3m_6+m_6^2)A+(m_2m_5+m_3m_4+m_1m_6)B \\
	       &+(m_1m_4+m_3m_5+m_2m_6)C+m_3m_6D+(m_1m_3+m_2m_5+m_4m_6)E \\
	       &+(m_2m_3+m_1m_4+m_5m_6)F+(m_1m_6+m_2^2+m_1m_3+m_3m_4+m_4m_6+m_5^2)G\\
	       &+(m_1^2+m_2m_6+m_2m_3+m_3m_5+m_4^2+m_5m_6)H \\
	       &+(m_1m_2+m_1m_5+m_2m_4+m_2m_4+m_1m_5 +m_4m_5)I+(m_1m_2+m_4m_5)J.
\end{aligned}
\end{equation*}
\end{theorem}
$\mathbf{Proof.}$
The result can be obtained directly from Lemma \ref{lemma 5}. $\blacksquare$

\section{Sextic residue double circulant self-dual codes }\label{section 3}

\subsection{General sextic residue double circulant self-dual codes}
\begin{theorem}\label{theorem 8}
Suppose $p=12l+7$ is an odd prime. Let $\alpha \in \mathrm{GF}(q)$, and $\vec{m} \in \mathrm{GF}(q)^7$. Then

(a) the code with the generator matrix $P_p(\vec{m})$ is a self-dual code on $\mathrm{GF}(q)$ if and only if the following holds
\begin{equation*}
D_0(\vec{m})=-1, D_1(\vec{m})=D_2(\vec{m})=D_3(\vec{m})=0.
\end{equation*}

(b) the code with the generator matrix $B_p(\alpha,\vec{m})$ is a self-dual code on $\mathrm{GF}(q)$ if and only if the following holds
\begin{equation*}
\begin{aligned}
\alpha^2+p=-1,\,\,\,  &-\alpha+m_0+f(m_1+m_2+m_3+m_4+m_5+m_6)=0, \\
D_0(\vec{m})=-2,\,\,\, &D_1(\vec{m})=D_2(\vec{m})=D_3(\vec{m})=-1.
\end{aligned}
\end{equation*}
\end{theorem}
$\mathbf{Proof.}$
According Theorem \ref{theorem 6}, we have
\begin{equation*}
\begin{aligned}
P_p(\vec{m})P_p(\vec{m})^T=&A_0+\sum_{i=0}^{6}D_i(\vec{m})A_i.
\end{aligned}
\end{equation*}
The following can be obtained according to the definitions of self-dual codes.
\begin{equation*}
B_p(\alpha,\vec{m})B_p(\alpha,\vec{m})^T=\begin{pmatrix}I_{p+1} & K\end{pmatrix}\begin{pmatrix}I_{p+1} \\ K^T\end{pmatrix}=I_{p+1}+KK^T,
\end{equation*}
where
\begin{equation*}
\begin{aligned}
KK^T&=
\left(
\begin{array}{cccccc}
\alpha &  1       &  \cdots  & 1 \\
-1     &  \quad   &  \quad   & \quad   \\
\vdots &  \quad   &  R       & \quad   \\
-1     &  \quad   &  \quad   & \quad
\end{array}
\right)
\left(
\begin{array}{cccccc}
\alpha &  -1       &  \cdots  & -1 \\
1     &  \quad   &  \quad   & \quad   \\
\vdots &  \quad   &  R^T       & \quad   \\
1     &  \quad   &  \quad   & \quad
\end{array}
\right) \\
&=
\left(
\begin{array}{cccccc}
\alpha^2+p &  S       &  \cdots  & S \\
S     &  \quad   &  \quad   & \quad   \\
\vdots &  \quad   &  X       & \quad   \\
S     &  \quad   &  \quad   & \quad
\end{array}
\right),
\end{aligned}
\end{equation*}
\begin{equation*}
\begin{aligned}
X=J_p+\sum_{i=0}^{6}D_i(\vec{m})A_i, S=-\alpha+m_0+\frac{p-1}{6}(m_1+m_2+m_3+m_4+m_5+m_6).
\end{aligned}
\end{equation*}
The result can be obtained from the definition of self-dual codes. $\blacksquare$

We now construct an infinite family of sextic residue double circulant self-dual codes on $\mathrm{GF}(q)$. In preparation, we have the following theorem.
\begin{theorem}\label{theorem 9}
Let p is an odd prime and $p=24k+19$, $k$ is a nonnegative integer. Suppose $g$ is the primitive root of $p$ ,and satisfies $g^m \equiv 2 \bmod p$. If $p=x^2+3y^2$, $x \equiv 1\bmod 3$, $y \equiv -m \bmod p$. Especially, we can reach one of the following equations
\begin{equation*}
\begin{aligned}
&(a) \qquad A\equiv B \equiv C \equiv E \equiv F \equiv 0 \bmod 2 ,D \equiv 1 \bmod 2,I \equiv J \equiv 1 \bmod 2,  G+H \equiv 1 \bmod 2, \\
&(b) \qquad A \equiv C \equiv D \equiv E \equiv F \equiv J \equiv 0 \bmod 2 ,B \equiv I \equiv 1 \bmod 2, G+H \equiv 1 \bmod 2, \\
&(c) \qquad A \equiv B \equiv C \equiv D \equiv E   \equiv J \equiv 0 \bmod 2 ,F \equiv I \equiv 1 \bmod 2, G+H \equiv 1 \bmod 2. \\
\end{aligned}
\end{equation*}
\end{theorem}
$\mathbf{Proof.}$
According Theorem \ref{theorem 4}, when $y=1 \bmod 3$, let $x=3a+1$, $y=3b+1$. Then
\begin{equation}
p=12l+7=x^2+3y^2=9a^2+1+6a+3(b^2+1+6b) \label{equation4-1}
\end{equation}
From Lemma \ref{theorem 4}, we know $B \equiv E$. Then£º
\begin{equation}
3(B+E)=\frac{48k+38+2-4(3a+1)-24(3b+1)}{12}=4k+1-a-6b \equiv 0 \bmod 2. \label{equation4-2}
\end{equation}
From Equation (\ref{equation4-2}), we have $a+1\equiv0\bmod 2$. And  from Equation (\ref{equation4-1}), we have $a^2+b^2\equiv 1 \bmod 2$. Thus $a$ is odd, while $b$ is even. Let $a=2c+1$, $b=2d$. So $x=6c+4$, $y=6d+1$. Then
\begin{equation*}
12l+7=(6c+4)^2+3(6d+1)^2.
\end{equation*}
Namely, $l\equiv (c+1) \bmod 2$.

Bring it into the cyclotomic numbers
\begin{equation*}
\left\{
\begin{aligned}	
&3A=\frac{12l+7-11-2(6c+4)}{12}=l-c-1\equiv 0\bmod 2, \\
&3B=\frac{12l+7+1-2(6c+4)-12(6d+1)}{12}=l-1-c-6d \equiv 0 \bmod 2, \\
&3C=\frac{12l+7+1-8(6c+4)+12(6d+1)}{12}=l-1-4c+6d \equiv 0 \bmod 2, \\
&3D=\frac{12l+7+1+10(6c+4)+12(6d+1)}{12}=l+5+5c+6d \equiv 0 \bmod 2, \\
&3F=\frac{12l+7+1+4(6c+4)}{12}=l+2c+2 \equiv 1 \bmod 2. \\
\end{aligned}
\right.
\end{equation*}
From Lemma \ref{theorem 4}, we know $I=F$, $J=C$, $B=E$.

When $y=2\bmod 3$, let $x=3a+1$, $y=3b+2$, we have
\begin{equation}
p=12l+7=x^2+3y^2=9a^2+1+6a+3(b^2+1+6b). \label{equation4-3}
\end{equation}
From Lemma \ref{theorem 4}, we know $C\equiv F$. Then
\begin{equation}
3(C+F)=\frac{48k+84-12a+72b}{12}=4k+7-a+6b\equiv 0\bmod 2. \label{equation4-4}
\end{equation}
From Equation (\ref{equation4-4}), we have $a+1\equiv0\bmod 2$. And from Equation (\ref{equation4-3}), we have $a^2+b^2\equiv 0 \bmod 2$. Thus $a$, $b$ are odd. Let $a=2c+1$, $b=2d+1$. So $x=6c+4$, $y=6d+5$. Then
\begin{equation*}
12l+7=9(2c+1)^2+1+6(2c+1)+3[9(2d+1)^2+1+6(2d+1)].
\end{equation*}
Namely, $l\equiv (c+1) \bmod 2$.

Bring it into the cyclotomic numbers
\begin{equation*}
\left\{
\begin{aligned}	
&3A=\frac{12l+7-11-2(6c+4)}{12}=l-c-1\equiv 0\bmod 2, \\
&3B=\frac{12l+7+1+4(6c+4)}{12}=l+2c+2 \equiv 1 \bmod 2, \\
&3C=\frac{12l+7+1-2(6c+4)+12(6d+5)}{12}=l+5-c+6d \equiv 0 \bmod 2, \\
&3D=\frac{12l+7+1+10(6c+4)-12(6d+5)}{12}=l-1+5c-6d \equiv 0 \bmod 2, \\
&3E=\frac{12l+7+1-8(6c+4)-12(6d+5)}{12}=l-7-4c-6d \equiv 0 \bmod 2, \\
&3J=\frac{12l+7+1-8(6c+4)+12(6d+5)}{12}=l-4c+6d+3 \equiv 0 \bmod 2.\\
\end{aligned}
\right.
\end{equation*}
From Lemma \ref{theorem 4}, we know $I=B$, $F=C$, $G+H\equiv 1\bmod2 $. \\
When $y=0\bmod 3$, let $x=3a+1$, $y=3b$. We have
\begin{equation}
p=12l+7=x^2+3y^2=9a^2+1+6a+9b^2. \label{equation4-5}
\end{equation}
From Lemma \ref{theorem 4}, we know $I \equiv J$. Then
\begin{equation}
3(I+J)=\frac{48k-12a+36b}{12}=4k-a+3\equiv 0\bmod 2. \label{equation4-6}
\end{equation}
From Equation (\ref{equation4-6}), we have $a+1\equiv0\bmod 2$. And  from Equation (\ref{equation4-5}), $a^2+b^2\equiv 0 \bmod 2$. Thus $a$, $b$ are odd. Let $a=2c+1$, $b=2d+1$. So $x=6c+4$, $y=6d+3$. Then
\begin{equation*}
12l+7=9(2c+1)^2+1+6(2c+1)+3[9(2d+1)^2].
\end{equation*}
Namely, $l\equiv (c+1) \bmod 2$.

Bring it into the cyclotomic numbers
\begin{equation*}
\left\{
\begin{aligned}	
&3A=\frac{12l+7-11-8(6c+4)}{12}=l-4c-3\equiv 0\bmod 2, \\
&3B=\frac{12l+7+1-2(6c+4)+12(6d+3)}{12}=l+3-c+6d \equiv 0 \bmod 2, \\
&3D=\frac{12l+7+1+16(6c+4)}{12}=l+6+8c \equiv 1 \bmod 2, \\
&3E=\frac{12l+7+1-2(6c+4)-12(6d+5)}{12}=l-5-c-6d \equiv 0\bmod 2,\\
&3I=\frac{12l+7+1-2(6c+4)}{12}=l-c \equiv 1 \bmod 2.
\end{aligned}
\right.
\end{equation*}
From Lemma \ref{theorem 4}, we know $B=C$, $E=F$, $I=J\equiv 1 \bmod 2$, and $G+H\equiv 1\bmod 2$. \ \ \ \ \ \ \ \ \ \ \ \ \ \ \ \ \ \ \ \ \ \ \ \ \ \ \ \ \ \ \ \ \ \ \ \ \ \ \ \ \ \ \ \ \ \ \ \ \ \ \ \ \ \ \ \ \ \ \ \ \ \ \ \ \ \ \ \ \ \ \ \ \ \ \ \ \ \ \ \ \ \ \ \ \ \ \ \ \ \ \ \ \ \ \ \ \ \ \ \ \ \ \ \ \ \ \ \ \ \ \ \ \ \ \ \ \ \ \ \ \ \ \ \ \ \ \ \ \ \ \ \ \ \ \ \ \ \ \ \ $\blacksquare$

\subsection{Sextic residue double circulant self-dual codes on $\mathrm{GF}(2)$}
\begin{theorem}\label{theorem 10}
Suppose $p$ is an odd prime and $p=24k+19$. Then we can get a  sextic residue pure double circulant  self-dual code on $\mathrm{GF(2)}$ with the generation matrix $P_p(\vec{m})$. And in certain conditions, it can be optimal.
\end{theorem}
$\mathbf{Proof.}$
If $p=24k+19$, from Theorem \ref{theorem 9}, let the first parameters as an example. We have
\begin{equation*}
\left\{
\begin{aligned}	
D_0(0,0,0,0,1,1,1)&\equiv 1\bmod 2, \\
D_1(0,0,0,0,1,1,1)&=A+B+C+2G+2H+I+J\equiv 0\bmod 2, \\
D_2(0,0,0,0,1,1,1)&=A+B+F+2G+2H+2I+2J\equiv 0\bmod 2, \\
D_3(0,0,0,0,1,1,1)&=A+E+F+2G+2H+I+J\equiv 0\bmod 2. \\
\end{aligned}
\right.
\end{equation*}
From Theorem \ref{theorem 8}, we know the generation matrix $P_p(\vec{m})$ on $\mathrm{GF(2)}$ is a self-dual code with length of $2p$. And the experimental results shows that the minimum Hamming distance is 8. So the codes whose parameters in Table \ref{table1} are  optimal codes according the list in \cite{Gaborit2020Tables}.$\blacksquare$
\begin{theorem}\label{theorem 11}
Suppose $p$ is an odd prime and $p=24k+19$. Then we can get a a  sextic residue double circulant  self-dual code with boundary on $\mathrm{GF(2)}$ which length is $2p+2$ with generation matrix $B_p(\alpha,m_0,m_1,m_2,m_3,m_4,m_5,m_6)$.  And in certain conditions, it can be optimal. \\
\begin{minipage}[t]{0.45\textwidth}
\centering
\tiny
\makeatletter\def\@captype{table}\makeatother\caption{\bf {Parameters of the code with $P_{41}$ }}\label{table1}
\begin{tabular}{|c|c|c|c|c|c|c|c|}
\hline$m_0$ & $m_1$ & $m_2$ & $m_3$ & $m_4$ & $m_5$ & $m_6$ & $MHD$ \\
\hline 0  & 0 & 0  & 0 & 1 & 1 & 1 & 8   \\
\hline 0  & 0 & 1  & 0 & 0 & 1 & 1 & 8   \\
\hline 0  & 0 & 1  & 0 & 1 & 0 & 1 & 8   \\
\hline 0  & 0 & 1  & 0 & 1 & 1 & 0 & 8   \\
\hline 0  & 0 & 1  & 1 & 0 & 1 & 0 & 8   \\
\hline 0  & 1 & 1  & 0 & 0 & 1 & 0 & 8   \\
\hline 0  & 1 & 1  & 1 & 0 & 0 & 0 & 8   \\
\hline 0  & 0 & 1  & 0 & 1 & 0 & 1 & 8   \\
\hline 0  & 0 & 1  & 1 & 1 & 0 & 0 & 8   \\
\hline 1  & 0 & 0  & 0 & 1 & 0 & 1 & 8   \\
\hline 0  & 1 & 0  & 0 & 0 & 1 & 1 & 8   \\
\hline 0  & 1 & 0  & 1 & 0 & 1 & 0 & 8   \\
\hline 0  & 1 & 1  & 0 & 0 & 0 & 1 & 8   \\
\hline 0  & 1 & 1  & 1 & 0 & 0 & 0 & 8   \\
\hline 1  & 0 & 1  & 1 & 1 & 1 & 0 & 8   \\
\hline 1  & 1 & 0  & 1 & 0 & 0 & 0 & 8   \\
\hline 0  & 0 & 1  & 0 & 1 & 0 & 1 & 8   \\
\hline
\end{tabular}
\end{minipage}
\begin{minipage}[t]{0.45\textwidth}
\renewcommand\arraystretch{1.575}
\centering
\tiny
\makeatletter\def\@captype{table}\makeatother\caption{\bf{Parameters of the code with $B_{41}$ }}\label{table2}
\begin{tabular}{|c|c|c|c|c|c|c|c|}
\hline$m_0$ & $m_1$ & $m_2$ & $m_3$ & $m_4$ & $m_5$ & $m_6$ & $MHD$ \\
\hline 0  & 0 & 0  & 1 & 1 & 1 & 1 & 8   \\
\hline 0  & 0 & 1  & 0 & 1 & 1 & 1 & 8   \\
\hline 0  & 0 & 1  & 1 & 1 & 0 & 1 & 8   \\
\hline 0  & 1 & 0  & 0 & 1 & 1 & 1 & 8   \\
\hline 0  & 1 & 0  & 1 & 0 & 1 & 1 & 8   \\
\hline 0  & 1 & 0  & 1 & 1 & 1 & 0 & 8   \\
\hline 0  & 1 & 1  & 0 & 1 & 0 & 1 & 8   \\
\hline 0  & 1 & 1  & 1 & 0 & 0 & 1 & 8   \\
\hline 0  & 1 & 1  & 1 & 0 & 1 & 0 & 8   \\
\hline 0  & 1 & 1  & 1 & 1 & 0 & 0 & 8   \\
\hline 1  & 0 & 0  & 0 & 1 & 1 & 1 & 8   \\
\hline 1  & 0 & 1  & 0 & 1 & 0 & 1 & 8   \\
\hline 1  & 1 & 0  & 1 & 0 & 1 & 0 & 8   \\
\hline 1  & 1 & 1  & 1 & 0 & 0 & 0 & 8   \\
\hline
\end{tabular} \\
\end{minipage}
\end{theorem}
$\mathbf{Proof.}$
If $p=24k+19$, from Theorem \ref{theorem 9}, let the first parameters as an example. We have
\begin{equation*}
\left\{
\begin{aligned}
D_0(0,0,0,1,1,1,1)&\equiv 0\bmod 2,\\
D_1(0,0,0,1,1,1,1)&=A+2B+C+E+F+3G+3H+3I+J\equiv 1\bmod 2, \\
D_2(0,0,0,1,1,1,1)&=A+B+C+E+2F+G+H+I+J\equiv 1\bmod 2, \\
D_3(0,0,0,1,1,1,1)&=A+B+C+D+E+F+G+H+I+J\equiv 1\bmod 2, \\
\alpha^2+p=1\bmod 2,&\\
-\alpha+m_0+\frac{p-1}{6}(&m_1+m_2+m_3+m_4+m_5+m_6)=0\bmod 2. \\
\end{aligned}
\right.
\end{equation*}

From Theorem \ref{theorem 8}, we know the code with the generation matrix $B_p(\alpha,m_0,m_1,m_2,m_3,m_4,m_5,m_6)$ on $\mathrm{GF(2)}$ is a self-dual code with length of $2p+2$. And the experimental results show that the minimum Hamming distance is 8. So the codes whose parameters are in Table \ref{table2} are optimal codes from the list in \cite{Gaborit2020Tables}. $\blacksquare$
\subsection{Sextic residue double circulant self-dual codes on $\mathrm{GF}(4)$}

Let $\varepsilon$ satisfy $\varepsilon^2+\varepsilon+1=0$ and be a fixed primitive element of $\mathrm{GF(4)}$. Then we have
\begin{theorem}\label{theorem 12}
Suppose $p=24k+19$ is an odd prime. Then we can get a a  sextic residue pure double circulant self-dual code on $\mathrm{GF(4)}$  with generation matrix  $P_p(\varepsilon,\varepsilon^2,1,\varepsilon^2,0,0,\varepsilon)$.  And in certain conditions, it can be optimal or best than ever.
\end{theorem}
$\mathbf{Proof.}$
Suppose $p=24k+19$ is an odd prime. From Theorem \ref{theorem 9}, we have
\begin{equation*}
\begin{aligned}
D_0(\varepsilon,\varepsilon^2,1,\varepsilon^2,0,0,\varepsilon)=&(4k+3)(1+\varepsilon+1+\varepsilon^2)\equiv 1 \bmod 2, \\
D_1(\varepsilon,\varepsilon^2,1,\varepsilon^2,0,0,\varepsilon)=&(A+B+C+3G+H+3I+J)\varepsilon  \\
&+(2B+E+F+2G +2H+I+J+1)\equiv 0 \bmod 2,\\
D_2(\varepsilon,\varepsilon^2,1,\varepsilon^2,0,0,\varepsilon)=&(A+C+J)+\varepsilon^2B+E\varepsilon+(1+\varepsilon^2)F \\ &+(2\varepsilon+\varepsilon^2)(G+H)+(\varepsilon^2+\varepsilon)I+\varepsilon\equiv 0\bmod 2, \\
D_3(\varepsilon,\varepsilon^2,1,\varepsilon^2,0,0,\varepsilon)=&(2A+C+E+F+G+3H+I+J+1)\varepsilon \\
&+(2A+B+D+F+2G+H+I+J)\equiv 0\bmod 2. \\
\end{aligned}
\end{equation*}
From Theorem \ref{theorem 8}, we know the code generate by matrix $P_p(\varepsilon,\varepsilon^2,1,\varepsilon^2,0,0,\varepsilon)$ on $\mathrm{GF(4)}$ is a self-dual code with length of $2p$.\ \ \ \ \ \ \ \ \ \ \ \ \ \ \ \ \ \ \ \ \ \ \ \ \ \ \ \ \ \ \ \ \ \ \ \ \ \ \ \ \ \ \ \ \ \ \ \ \ \ \ \ \ \ \ \ \ \ \ \ \ \ \ \ \ \ \ \ \ \ \ \ \ \ \ \ \ \ \ \ \ \ \ \ \ \ \ \ \ \ \ \ \ \ \ \ \ \ \ \ \ \ \ \ \ \ \ \ \ \ \ \ \ \ \ \ \ \ $\blacksquare$

 As an example,  Table \ref{table3} gives some $[38,19,11]_4$ linear codes  by Theorem \ref{theorem 12} which can reach the lower bound in Lemma \ref{Lemma 2.1} or possess the best than ever bound \cite{Gaborit2020Tables}. \\
\begin{minipage}[t]{0.45\textwidth}
\renewcommand\arraystretch{1.54}
\centering
\tiny
\makeatletter\def\@captype{table}\makeatother\caption{\bf{Parameters of codes with $P_{19}$}}\label{table3}
\begin{tabular}{|c|c|}
\hline$construction$ & $Remark$ \\
\hline $P_{19}(\varepsilon^2,\varepsilon^2,\varepsilon,1,\varepsilon,0,0)$   & Best than ever   \\
\hline $P_{19}(\varepsilon,\varepsilon^2,0,0,\varepsilon,\varepsilon^2,1)$   & Best than ever     \\
\hline $P_{19}(\varepsilon,\varepsilon^2,1,\varepsilon^2,0,0,\varepsilon)$   & Best than ever     \\
\hline $P_{19}(\varepsilon,\varepsilon^2,\varepsilon,1,\varepsilon^2,1,1)$   & Optimal code     \\
\hline $P_{19}(\varepsilon^2,\varepsilon^2,1,\varepsilon,\varepsilon^2,\varepsilon,\varepsilon^2)$   & Optimal code     \\
\hline
\end{tabular} \\
\end{minipage}
\begin{minipage}[t]{0.45\textwidth}
\centering
\tiny
\makeatletter\def\@captype{table}\makeatother\caption{\bf{Parameters of codes with $B_{19}$}}\label{table4}
\begin{tabular}{|c|c|}
\hline$construction$ & $Remark$ \\
\hline $B_{19}(0,0,1,1,\varepsilon,\varepsilon^2,\varepsilon^2,\varepsilon)$   & Optimal code   \\
\hline $B_{19}(0,\varepsilon,\varepsilon,\varepsilon^2,\varepsilon,\varepsilon^2,0)$   & Optimal code      \\
\hline $B_{19}(0,0,\varepsilon,\varepsilon,\varepsilon,\varepsilon^2,\varepsilon,0)$   & Optimal code     \\
\hline $B_{19}(0,0,\varepsilon,\varepsilon^2,\varepsilon^2,\varepsilon,1,1)$   & Optimal code     \\
\hline $B_{19}(0,\varepsilon,\varepsilon^2,\varepsilon,\varepsilon,\varepsilon,1,\varepsilon)$   & Optimal code      \\
\hline $B_{19}(0,0,\varepsilon^2,\varepsilon^2,\varepsilon,1,1,\varepsilon)$   & Optimal code \\
\hline
\end{tabular} \\
\end{minipage} \\
\begin{theorem}\label{theorem 13}

Suppose $p=24k+19$ is an odd prime. Then we can get a a  sextic residue double circulant  self-dual code  with boundary on $\mathrm{GF(4)}$  with generation matrix  $B_p(0,0,1,1,\varepsilon,\varepsilon^2,\varepsilon^2,\varepsilon)$.  And in certain conditions, it can be optimal.
\end{theorem}
$\mathbf{Proof.}$
Let $p=24k+19$ is an odd prime, then from Theorem \ref{theorem 9},
\begin{equation*}
\left\{
\begin{aligned}
D_0(0,0,1,1,\varepsilon,\varepsilon^2,\varepsilon^2,\varepsilon)&=(4k+3)(1+\varepsilon+\varepsilon^2)\equiv 0\bmod 2, \\
D_1(0,0,1,1,\varepsilon,\varepsilon^2,\varepsilon^2,\varepsilon)&=(D+E+G+H+I+J)\varepsilon+(D+E+G+H+I+J)\equiv 1\bmod 2, \\
D_2(0,0,1,1,\varepsilon,\varepsilon^2,\varepsilon^2,\varepsilon)&=(C+D+G+H+I+J)\varepsilon+(C+D+G+H+I+J)\equiv 1\bmod 2, \\
D_3(0,0,1,1,\varepsilon,\varepsilon^2,\varepsilon^2,\varepsilon)&=(A+D+G+H+I+J)\varepsilon+(A+D+G+H+I+J)\equiv 1\bmod 2, \\
\alpha^2+p=1\bmod 2, \,\,\,\,\,\,\,& \\
-\alpha+m_0+\frac{p-1}{6}(m_1&+m_2+m_3+m_4+m_5+m_6)=0\bmod 2. \\
\end{aligned}
\right.
\end{equation*}
From Theorem \ref{theorem 8}, we know that the code with the generation matrix $B_p(\alpha,m_0,m_1,m_2,m_3,m_4,m_5,m_6)$ on $\mathrm{GF(4)}$ is a self-dual code with length of $2p+2$.  $\blacksquare$

As an example,  Table \ref{table4} gives some $[40,20,12]_4$ linear codes constructed by Theorem \ref{theorem 13} which can reach the lower bound in Lemma \ref{Lemma 2.1}.

\section{Summary}
In this paper, we construct sextic residue pure double circulant codes and sextic residue double circulant codes with boundary from sextic cyclotomy, and give the necessary and sufficient conditions for them to be self-dual. Numerical experiments show that these self-dual codes contain some optimal ones and best than ever ones on $\mathrm{GF}(2)$ and $\mathrm{GF}(4)$. We believe that the construction method on residues of higher degrees and the ones on the generalized cyclotomy of rings are rich sources of self-dual codes with good parameters.

\section{Acknowledgement}
This work was supported by Fundamental Research Funds for the Central Universities (No. ZD2019-183-008), the Major Scientific and Technological Projects of CNPC under Grant ZD2019-18 (No. ZD2019-183-001), and Shandong Provincial Natural Science Foundation of China (ZR2019MF070).
\bibliography{bibfile}
\end{document}